\documentstyle[12pt]{article}

\begin{document} 
\begin{titlepage}
\today          \hfill
\begin{center}
\hfill    LBNL-41755 \\
\hfill    UCB-PTH-98/21

\vskip .05in

{\large \bf An Improved Model of Direct Gauge Mediation}
\footnote{This work was supported in part
by the Director, Office of Energy
Research, Office of High Energy and Nuclear Physics, Division of High
Energy Physics of the U.S. Department of Energy under Contract
DE-AC03-76SF00098 and in part by the National Science Foundation
under grant PHY-90-21139, and also by the Berkeley Graduate Fellowship.}
\vskip .15in
Kaustubh Agashe \footnote{email: KSAgashe@lbl.gov}\\
{\em
    Theoretical Physics Group\\
    E. O. Lawrence Berkeley National Laboratory\\}
and \\
{\em
    Department of Physics \\
    University of California\\
    Berkeley, California 94720}
\end{center}

\vskip .05in

\begin{abstract}
We present a new, 
improved model of gauge mediation 
of dynamical SUSY Breaking:
the model does not have gauge messengers or $\sim 10$ TeV                       
scalars charged under the Standard Model (SM), thus                             
avoiding the problem of
negative (mass)$^2$ for supersymmetric SM (SSM)
scalars faced by some earlier models.
The gauge mediation is 
direct, {\it i.e.,}
the messengers which communicate SUSY breaking to the SSM fields
carry quantum numbers of the gauge group which
breaks SUSY.
These messenger fields couple to a modulus field.
The model has a very simple particle content:
the modulus and the messengers are the only
chiral superfields (other than the SSM fields) in the model.
The inverted hierarchy mechanism is used to 
generate a local SUSY breaking minimum
for the modulus field in a perturbative regime thus making 
the model calculable. 
\end{abstract}


\end{titlepage}
\renewcommand{\thepage}{\roman{page}}
\setcounter{page}{2}
\mbox{ }

\vskip 1in

\begin{center}
{\bf Disclaimer}
\end{center}

\vskip .2in

\begin{scriptsize}
\begin{quotation}
This document was prepared as an account of work sponsored by the United
States Government. While this document is believed to contain correct
 information, neither the United States Government nor any agency
thereof, nor The Regents of the University of California, nor any of their
employees, makes any warranty, express or implied, or assumes any legal
liability or responsibility for the accuracy, completeness, or usefulness
of any information, apparatus, product, or process disclosed, or represents
that its use would not infringe privately owned rights.  Reference herein
to any specific commercial products process, or service by its trade name,
trademark, manufacturer, or otherwise, does not necessarily constitute or
imply its endorsement, recommendation, or favoring by the United States
Government or any agency thereof, or The Regents of the University of
California.  The views and opinions of authors expressed herein do not
necessarily state or reflect those of the United States Government or any
agency thereof, or The Regents of the University of California.
\end{quotation}
\end{scriptsize}

\vskip 2in

\begin{center}
\begin{small}
{\it Lawrence Berkeley Laboratory is an equal opportunity employer.}
\end{small}
\end{center}

\newpage
\renewcommand{\thepage}{\arabic{page}}
\setcounter{page}{1}
\section{Introduction}
There has been a substantial effort in the last few years
in building models of gauge mediated dynamical SUSY 
breaking (for a review, see \cite{giudice1}). 
These models are predictive since they have only a few
parameters and
solve the supersymmetric flavor problem:
the supersymmetric contributions to the
flavor changing neutral currents (FCNC's) are   
negligible since the scalars with the same gauge quantum numbers
are degenerate. Typically, these models contain
vector-like fields with Standard Model (SM)
gauge quantum numbers, called ``messengers",
which couple to a flat direction
(modulus) of the model. This modulus develops 
a vacuum expectation value (vev) in it's scalar and
$F$ components resulting in a non-supersymmetric
spectrum for the messengers.
The messenger fields then communicate SUSY
breaking to
the supersymmetric SM (SSM) sparticles at one/two loops.
Dine {\it et al.}
first constructed models in which the 
messengers 
are {\it not} a part of
the dynamical SUSY breaking (DSB) sector\cite{dine}. Recently, models
with the messengers as an integral part of DSB sector ({\it i.e.}
carrying the quantum numbers of the gauge group which
breaks SUSY) have been 
built\cite{poppitz1,nima1,hitoshi,dim1,dim2,luty,shirman,terning}.
We will call these
{\it direct} gauge mediation (GM) models.
We refer the reader to the review by Giudice and Rattazzi \cite{giudice1}
for other models of gauge mediation. 

We briefly mention the mechanisms used to stabilize the modulus
and break SUSY in some of the 
existing models of direct GM and also the problems faced by 
some of these models.

In the models of references \cite{poppitz1}
and \cite{nima1}, a dynamical superpotential leads to a
run-away behaviour along a classical flat direction. Non-renormalizable
operators stabilize the potential at large expectation values.
In the model of reference \cite{poppitz1}, 
the SUSY breaking scale is so high that
the Supergravity
contributions to the scalar soft masses dominate over the GM 
contributions possibly leading to non-degenerate soft masses.
This problem was overcome in the model of reference \cite{nima1}.
The models of both references \cite{poppitz1} 
and \cite{nima1} have scalars charged under the SM originating
in the DSB sector with soft masses $\sim 10$ TeV. These scalars
drive the SSM scalar (mass)$^2$ negative through two loop
Renormalization Group Evolution (RGE) \cite{nima1,poppitz2}.

References \cite{hitoshi,dim1} use the ``inverted hierarchy mechanism"
(loop
corrections) to
generate a SUSY breaking minimum (which might be a local minimum)
along a direction
with a constant non-zero potential energy at tree level.
These models have massive gauge multiplets with a non supersymmetric
spectrum. The SSM scalars get a negative contribution to their
soft (mass)$^2$ by coupling at two loops to these heavy gauge
multiplets (we will refer to them as ``gauge" messengers)\cite{giudice2}.
To avoid this problem, the authors of \cite{dim2}
used a singlet as the modulus and, to realize
the inverted hierarchy, had to add another singlet and some extra
matter fields.

A new mechanism of SUSY breaking, which could be used
for direct gauge mediation,  was discussed in \cite{terning}.
In this model, different mechanisms lift the flat directions in
different regions of the classical moduli space.
This model requires
a dynamical assumption (about
non-calculable terms in the K\"ahler potential) to work.

In this 
letter, we present a new, improved model of direct gauge mediation: 
there are no gauge messengers or $\sim 10$ TeV scalars charged under the SM,
thus avoiding the problems of negative SSM scalar (mass)$^2$
faced by the models in \cite{poppitz1,nima1,hitoshi,dim1}.
The model has 
a very simple
structure and particle content: the only chiral superfields
(in addition to the SSM fields) in the model are the modulus and
the messengers,
unlike the model in \cite{dim2}.
The model 
uses the inverted hierarchy
mechanism to stabilize the vev of the 
modulus in a perturbative regime so that the model is calculable.
\section{The Model}
The gauge group and global symmetry group of the model are
\begin{equation}
SU(2)_1 \times SU(2)_2 \times [ SU(6) \times U(1)_R ],
\end{equation}
where the group in brackets is the global symmetry group.
Later, we will identify part of this global symmetry with the SM gauge group.
The particle content is
\begin{eqnarray}
\Sigma
& \sim
\; ({\bf 2},{\bf 2}) & \times ({\bf 1},2) \nonumber \\
Q
& \sim
\; ({\bf 2},{\bf 1}) & \times ({\bf 6},0) \nonumber \\
\bar{Q}
& \sim
\; ({\bf 1},{\bf 2}) & \times (\overline{{\bf 6}},0).
\end{eqnarray}
The only renormalizable
superpotential consistent with the gauge and global
symmetries is
\begin{equation}
W = \lambda \Sigma Q \bar{Q},
\label{W}
\end{equation}
where the gauge and global indices are appropriately summed over.
The non anomalous $U(1)_R$ symmetry was imposed to forbid mass terms for 
the fields. With the superpotential in Equation \ref{W}, this
is the only non-anomalous $U(1)$ symmetry of the model.
Consider the $D$-flat direction parametrized by $\hbox{det} \Sigma$. 
Up to global and gauge symmetries, the $\Sigma$ vev along this 
direction is 
\begin{equation}
\langle \Sigma \rangle = \frac{1}{\sqrt{2}}\hbox{diag}[v,v].
\end{equation}
This vev breaks $SU(2)_1 \times SU(2)_2$ to the diagonal 
$SU(2) _D$ at the scale $v/\sqrt{2}$.
Three components of $\Sigma$ are eaten by the
super-Higgs mechanism. The remaining 
component (the superfield $1/\sqrt{2} \;\hbox{tr}\Sigma$)
which is the flat direction is massless and is a singlet of
$SU(2)_D$. 
We denote this superfield (and the vev of it's
scalar component) by $v$.
Assume
$v >> \Lambda _1, \Lambda _2$ where
the $\Lambda _1, \Lambda _2$
are the dynamical scales of the $SU(2)$ groups
so that the gauge couplings are weak at the scale
$v$ and it suffices to use
tree level
matching and one loop running of
gauge couplings.\footnote{
However, see the second footnote on page 6.}  
We  
match the holomorphic gauge couplings of $SU(2)_{1,2}$ and $SU(2)_D$
at the scale $v/\sqrt{2}$: $1/g_D^2 (v/\sqrt{2}) = 
1/g_1^2 (v/\sqrt{2}) + 1/g_2^2 (v/\sqrt{2})$ (this is the first
matching condition).\footnote{The 
canonical gauge couplings
should be matched at the mass of the heavy gauge boson
$\sim \sqrt{g_1^2 + g_2^2} \; v/\sqrt{2}$. Using the
Shifman-Vainshtein formula\cite{shifman} for the relation
between the canonical and holomorphic
gauge couplings, we can show that the holomorphic gauge couplings
have to be matched at the scale $v/\sqrt{2}$\cite{nima2}.} 
The diagonal $SU(2) _D$ has 12 fundamentals, $Q, \bar{Q}$
and thus, at one loop, it's gauge coupling does not run
between the scale $v/\sqrt{2}$ and the scale $\lambda
/ \sqrt{2}
\; v$
where
all of the $Q, \bar{Q}$ become heavy (assume, for simplicity, that
$\lambda 
< 1$).
Below the scale $\lambda v/\sqrt{2}$, we then 
have a pure $SU(2)$ gauge theory with the
singlet superfield $v$. 
At the scale $\lambda v /\sqrt{2}$
where $Q,\bar{Q}$ are integrated out, 
we set the gauge 
coupling of the $SU(2)_D$ theory
with $Q,\bar{Q}$
equal to the gauge coupling of
the pure $SU(2)$ gauge theory (this is the second matching condition).
The two matching conditions
give the holomorphic dynamical scale, $\Lambda _L$,
of the pure $SU(2)_D$:
\begin{equation}
\label{matching}
\left( \frac{\Lambda _L}{\lambda v / \sqrt{2}} \right) ^6 = \left( 
\frac{\Lambda_1}{v / \sqrt{2}} \right) ^ 2 \left(
\frac{\Lambda_2}{v / \sqrt{2}} \right) ^ 2.
\end{equation}
This theory undergoes gaugino condensation giving the
superpotential\footnote{A priori, 
there could be a superpotential
term induced by the instantons in the broken $SU(2)$ gauge group
(even though the
$SU(2) \times SU(2)$ gauge group
is not completely broken)
\cite{csaki}. However, along the flat direction det$\Sigma$
(for $v >> \Lambda _1, \Lambda _2$),
Equation \ref{exact} is the only superpotential consistent with
the non-anomalous $R$-symmetry and 
anomalous $U(1)$ symmetries acting on 
$Q$ and $\bar{Q}$
and thus no other superpotential term can
be generated.}:
\begin{eqnarray}
W_{eff} & = & 2\; \Lambda _L^3 \nonumber \\
 & = & \sqrt{2}\; \lambda ^3
\Lambda _1 \Lambda _2 v. 
\label{exact}
\end{eqnarray}
The low energy theory (below $\Lambda _L$)
has only the field $v$ and has $F_v = \sqrt{2} \; \lambda ^3 \Lambda ^2$
where $\Lambda ^2 = \Lambda _1 \Lambda _2$. 
At tree level, the superfield $v$ has
the canonical K\"ahler potential $v^{\dagger}v$ in the low energy theory.
Thus, the model breaks SUSY
with a constant vacuum energy $2 \; \lambda ^6
\Lambda ^4$.
The vev $v$ is undetermined at this level. To determine $v$, we need to include the corrections to the K\"ahler
potential of $v$ due to the wavefunction renormalization $Z$ for
$\Sigma$\cite{hitoshi,dim1,nima2}. This is the only modification
to the potential since the superpotential in
Equation \ref{exact} is exact\cite{nima2}. 
The effective low energy Lagrangian is:
\begin{equation}
{\cal L} = \int d^4 \theta Z(v) v^{\dagger}v + \int d^2 \theta \sqrt{2} \;
\lambda^3
\Lambda ^2 v + \hbox{h.c.}.
\end{equation}
(There is no renormalization 
of $Z$ below the scale $v$ since all the fields coupling
to $\Sigma$ become heavy at $\sim v$.)
The potential is then\cite{hitoshi,dim1,nima2}
\begin{equation}
V(v) = \frac{2 \; \lambda ^6 \Lambda ^4}{Z(v)}.
\end{equation}
Since $v >> \Lambda _1, \Lambda _2$, we can compute $Z$
in perturbation theory.
The one loop RGE for $Z$ is
\begin{equation}
\frac{d Z(t)}{d t} = \frac{2 Z(t)}
{16 \pi ^2} \left( \frac{3}{2}(g_1(t)^2 + g_2(t)^2)
- 6 \lambda (t)^2 \right).
\label{Z}
\end{equation}
where $g_1$ and $g_2$ are the gauge couplings of the two $SU(2)$ gauge groups
and $t \sim \ln v$.
The potential develops a minimum along $v$ via the
``inverted hierarchy mechanism''\cite{witten} as follows.
At large momentum scales, it is possible that the Yukawa
contribution in the above equation
dominates, since the gauge couplings are asymptotically
free while the Yukawa coupling can grow with energy. This makes
$Z$ smaller as $v$ increases. Similarly, for small values of $v$,
the gauge contribution to Equation \ref{Z} dominates. Thus, for small $v$,
$Z$ increases with energy. In other words, the potential decreases
with energy for small $v$ and increases with $v$ for large $v$. So,
we get a minimum of the potential for $v$ such that
$\lambda \sim g$ so that $dZ/dt$ is zero.
This scale can be naturally much larger than $\Lambda_{1,2}$
since the RG scaling is logrithmic in $v$. We need $v >> \Lambda_{1,2}$
so that we are in the perturbative regime. 
Upon gauging the $SU(3) \times SU(2) \times U(1)$ subgroup of the
global $SU(6)$ symmetry,  
the $Q,\bar{Q}$ act as two $5 + \bar{5}$ messengers since 
they have a supersymmetric mass $\lambda v/\sqrt{2}$ and a SUSY breaking
(mass)$^2$, 
$\lambda F_v \sqrt{2}\; Q \bar{Q}$ (here $Q$ and
$\bar{Q}$ denote the scalar components).
Thus the SM
gauge couplings remain perturbative upto the GUT scale
even for small values of $v$.
We need
$F_v / v = \sqrt{2} \; \lambda ^3 \Lambda ^2 /v \sim 10 -100$ TeV to get
SSM scalar and gaugino masses in the range $100$ GeV - $1$ TeV.

The scalar contained in the superfield $v$ acquires a mass 
$\sim F_v/v 
\times (1/16 \pi ^2)$ 
$\sim \hbox{few}\; 100$ GeV (for $\lambda, \; g \sim 1$)
once the local
minimum develops. This scalar is not charged under the SM.
The spontaneous breaking of the $U(1)_R$ symmetry produces a 
Nambu-Goldstone boson which consists mainly of the pseudoscalar in $v$.
This ``R-axion'' can acquire a mass greater than about $100$ MeV
if we add a constant term to the superpotential to cancel the 
cosmological constant once SUSY is made local\cite{poppitz3}.
Then, the R-axion
is safe from astrophysical constraints.
The fermion in $v$ is the Goldstino and is eaten by the
gravitino when SUSY is made local.
This model has no gauge messengers: the broken $SU(2)$ 
gauge multiplet does have a non-supersymmetric spectrum, but 
it does not couple to the SSM scalars at one or two loops.
The phenomenology of this model is
similar to that of conventional gauge mediation with two families
of messengers and messenger scale $v$\cite{pheno}.

There are other flat directions, $Q^2$ and $\bar{Q}^2$ (with the
$SU(6)$ symmetry global).
Consider the flat direction $Q_1 . Q_2 \neq 0$.
Along this direction, $SU(2)_1$ is broken and
$\Sigma$ and two $\bar{Q}$ s
become massive.
The low energy theory is $SU(2)_2$ with four fundamentals $(\bar{Q})$
which has a moduli space with a quantum
modified constraint\cite{seiberg}. 
Thus, no superpotential is generated along the 
$Q^2$ flat direction\footnote{A priori, the instantons of the
completely broken $SU(2)_1$ group can generate a superpotential.
Any non-perturbative
superpotential has charge 2 under the non-anomalous
$U(1)_R$ symmetry. The only field with a non-zero charge 
under this symmetry, $\Sigma$, is heavy along this flat direction. 
Thus we expect no superpotential to
be generated along this flat direction.}.
The K\"ahler potential is known for $Q_1.Q_2 >> \Lambda _1^2, 
\Lambda _2^2$ (it is canonical in $Q$). 
Thus, there is a SUSY minimum along the $Q^2$
flat direction for $Q^2 >> \Lambda _1^2, \Lambda _2^2$.
A similar analysis is true
for the $\bar{Q}^2$ direction.

For
vev's $\sim O( \Lambda _1, \Lambda _2 )$
along the flat directions, the $SU(2)$'s are
strongly coupled (at the scale of the vevs)
and hence the above analysis is not valid.
For example, along the flat direction det$\Sigma$, for $v$
not much larger than $\Lambda _{1,2}$, we can still
integrate out $Q,\bar{Q}$ but 
there will be higher loop and non perturbative effects in the matching
of the dynamical scales, Equation \ref{matching}.
\footnote{Using the techniques of
\cite{nima3,michael}, {\it i.e.}, invariances under 
various $U(1)$ symmetries, it is possible to show that,
in this case, the matching of Equation \ref{matching} is exact
even non-perturbatively.}
Also, for vev's 
${\stackrel{<}{\sim}} O( \Lambda _1, \Lambda _2 )$, 
the fields $Q,\bar{Q}$ along
the flat direction det$\Sigma$ and similarly $\Sigma,\bar{Q}$
along the $Q^2$ flat direction can not be integrated out, {\it i.e.},
these fields appear as additional light degrees of freedom
in the superpotential.
The K\"ahler potential along the flat directions is also not
calculable for small vev's. We require a weakly coupled description
for this purpose.

Each $SU(2)$ considered separately has 
four flavors and thus has a dual description. 
In \cite{poppitz4}, dual theories to the $SU(2) \times SU(2)$  
models of this kind were constucted and it was shown that the
dual theories have the same infra-red physics as the original theories.
We checked that in these dual theories,
on adding the superpotential of Equation \ref{W},
the non-perturbative superpotential of Equation \ref{exact} is generated
and that along the flat direction
corresponding to $Q^2$ no superpotential
is generated (for vev's along the flat directions
such that we are in the perturbative regime). 
However, in these dual theories, there is always one gauge
group which is strongly coupled
in the infra-red so that the dynamics and the
K\"ahler potential for small vev's are still not calculable.
It is possible that there is a SUSY minimun near the origin (in addition
to the SUSY minimum along $Q^2 \; \hbox{or}\; \bar{Q}^2 
>> \Lambda _1^2, \Lambda _2^2$ with $\Sigma = 0$).

Since the model is non-chiral, {\it i.e.}, we can
add mass terms to all the fields (with the $SU(6)$ global), we do expect  
a global SUSY preserving minimum.	
Thus, the minimum we obtained along the $v$ direction is only
a local minimum. 
We can estimate the tunneling rate from this false vacuum
to the SUSY preserving minimum along the $Q^2 \; \hbox{or}\; 
\bar{Q}^2\neq 0,
\Sigma \sim 0$
flat directions. 
The potential energy, $E$, of the false vacuum is
$\sim \Lambda _1 ^2 \Lambda _2 ^2$. 
The distance in field space, $\Delta \Phi$,
from the false vacuum to the true vacuum
is $\sim v$. Since, $\Delta \Phi >> E^{1/4}$, the tunneling action 
from the ``false"
vacuum to the true vacuum 
can be estimated as \cite{dim1,luty}:
\begin{equation}
S \sim 2 \pi ^2 \frac{(\Delta \Phi)^4}{E} 
\sim 2 \pi^2 \frac{v^4}{\Lambda _1 ^2 \Lambda _2 ^2}. 
\end{equation}
Thus, the tunneling rate is negligibly small since  $v >> 
\Lambda _1, \Lambda _2$
which was required for a perturbative calculation.

We did a numercial analysis to find out the range
of possible values of $v$. We proceed as follows. We choose a
value of $v$ 
and choose $\Lambda$ such that
$\Lambda ^2 /v \sim 10 -100$ TeV to get
SSM scalar and gaugino masses in the range $100$ GeV - $1$ TeV. 
We assume that the wavefunction $Z$ is 1 at the Grand Unification
(GUT) scale.
Assuming 
$\Lambda _1 = \Lambda _2$, for simplicity, gives
the $SU(2)$ couplings at the GUT scale. 
When we gauge the SM subgroup of the global $SU(6)$ symmetry,
the Yukawa couplings for the different $SU(6)$ components of
$Q,\bar{Q}$ are no longer the same
at all energies due to RG scaling. For simplicity, we
assume that the $\lambda$'s are all equal at the GUT scale.
The value of $v$ along with the weak scale values of the SM 
gauge couplings gives us the SM gauge couplings at the GUT scale.
Then, with these boundary conditions (at the GUT scale),
we numerically solve the RGE's for $Z$ and $\lambda$'s
to determine
the value of $\lambda$ at the GUT scale which gives
a minimum at the chosen value of $v$. 
There is no solution for $\lambda$ if $v 
{\stackrel {<}{\sim}} 10^{10}$ GeV.
The reason is as follows. For $v {\stackrel {<}{\sim}}
10^{10}$ GeV and $\Lambda ^2 /v 
\sim 10 -100$ TeV, we get $v/\Lambda _{1,2} 
{\stackrel {<}{\sim}} 10^{3}$ which implies that
$g_{1,2} (v)$ and hence the $\lambda (v) \sim g_{1,2} (v)$ 
(required for a minimum at $v$) are $\sim 2 \;-\; 3$.  This results in
the Yukawa couplings hitting their Landau poles
below the GUT scale. 
We checked that it is possible to get
a consistent minimum for 
values of $v$ between $10^{10}$ GeV and $10^{15}$ GeV
for $O(1)$ values of $\lambda$ and the $SU(2)$ gauge coupling
at the GUT scale. For larger values of
$v$ and hence $F_v$, the supergravity contribution to the scalar
masses begins to dominate spoiling the degeneracy of the squarks and
sleptons. 

In summary, we have presented a simple model of direct 
gauge mediation which uses the inverted hierarchy mechanism to generate a
local SUSY breaking minimum.
The model does not have gauge messengers
or $\sim 10$ TeV scalars charged under the SM and thus it
avoids the problem of negative SSM scalar (mass)$^2$ faced by 
some of the earlier
models of direct GM. This model works for messenger scales
between $10^{10}$ GeV and $10^{15}$ GeV.

\section{Acknowledgements}
The author would like to thank Nima Arkani-Hamed, 
Csaba Cs\'aki, Michael Graesser,
Takeo Moroi, Hitoshi Murayama and John Terning for useful discussions
and Csaba Cs\'aki, 
Ian Hinchliffe, Takeo Moroi and Mahiko Suzuki for reading the manuscript.
The work was supported in part by the Director, Office
of Energy Research,
Office of High Energy Physics, Division of High Energy Physics of the
U.S. Department of Energy under Contract DE--AC03--76SF00098 and in part
by the National
Science Foundation under grant PHY-95-14797, and also by the Berkeley
Graduate Fellowship.


\begin{thebibliography}{99}
\bibitem{giudice1} G. Giudice, R. Rattazzi, hep-ph/9801271. 
\bibitem{dine} M. Dine, A.E. Nelson, Phys. Rev. {\bf D48}, 1277 (1993);
M. Dine, A.E. Nelson, Y. Shirman,
Phys. Rev. {\bf D51}, 1362 (1995); M. Dine, A.E. Nelson, Y. Nir, Y. Shirman,
Phys. Rev. {\bf D53}, 2658 (1996).
\bibitem{poppitz1} E. Poppitz, S.P. Trivedi, Phys. Rev. {\bf D55}, 5508 (1997).
\bibitem{nima1} N. Arkani-Hamed, H. Murayama, 
J. March-Russell, 
Nucl. Phys. {\bf B509}, 3 (1998).
\bibitem{hitoshi} H. Murayama,
Phys. Rev. Lett. {\bf 79}, 18 (1997).
\bibitem{dim1} S. Dimopoulos, G. Dvali, R. Rattazzi, G. Giudice,
Nucl. Phys. {\bf B510}, 12 (1998).
\bibitem{dim2} S. Dimopoulos, G. Dvali, R. Rattazzi,
Phys. Lett. {\bf B413}, 336 (1997).
\bibitem{luty} M.A. Luty,
Phys. Lett. {\bf B414}, 71 (1997).
\bibitem{shirman} Y. Shirman,
Phys. Lett. {\bf B417}, 281 (1998).
\bibitem{terning} M.A. Luty, J. Terning, hep-ph/9709306. 
\bibitem{poppitz2} E. Poppitz, S.P. Trivedi, 
Phys. Lett. {\bf B401}, 38 (1997).
\bibitem{giudice2} G. Giudice, R. Rattazzi, 
Nucl. Phys. {\bf B511}, 25 (1998).
\bibitem{shifman}M. A. Shifman, A. I. Vainshtein, Nucl. Phys.
{\bf B277}, 456 (1986).
\bibitem{nima2} N. Arkani-Hamed, H. Murayama, hep-th/9705189, 
to be published in Phys. Rev. {\bf D}.
\bibitem{csaki}C. Cs\'aki, H. Murayama, hep-th/9804061. 
\bibitem{witten} E. Witten, Phys. Lett. {\bf B105}, 267 (1981).
\bibitem{poppitz3} J. Bagger, E. Poppitz, L. Randall, Nucl. Phys.
{\bf B426}, 3 (1994).
\bibitem{pheno}see, for example: S. Dimopoulos, S. Thomas, J. Wells,
Nucl. Phys. {\bf B488}, 39 (1997).
For implications for cosmology and nucleosynthesis, see:
A. de Gouv\^ea, T. Moroi, H. Murayama 
Phys. Rev. {\bf D56}, 1281 (1997);
\cite{nima1};\cite{hitoshi};\cite{dim1}.
\bibitem{seiberg}N. Seiberg, Phys. Rev. {\bf D49}, 6857 (1994).
\bibitem{nima3}N. Arkani-Hamed, H. Murayama, hep-th/9707133.
\bibitem{michael}M. Graesser, B. Morariu, hep-th/9711054, to be published
in Phys. Lett. {\bf B}.
\bibitem{poppitz4} E. Poppitz, Y. Shadmi, S.P. Trivedi, 
Nucl. Phys. {\bf B480}, 125 (1996).
\end{thebibliography}
\end{document}